\documentclass[manuscript]{aastex631}

\usepackage[utf8]{inputenc}
\usepackage{graphicx}
\usepackage{natbib}
\usepackage{subfigure}
\usepackage{subfigure}
\usepackage{graphicx}
\usepackage{amsmath}
\usepackage{booktabs} 
\let\tablenum\relax 
\usepackage{siunitx}
\begin{document}

\title{Interferometric molecular line observations toward the 21\,$\mu$m protoplanetary nebula IRAS~$06530-0213$}

\shorttitle{Molecular line observations toward  IRAS~06530-0213}
\shortauthors{Sun et al.}

\correspondingauthor{Yong Zhang}
\email{zhangyong5@mail.sysu.edu.cn}

\author[0009-0007-4663-2643]{Hao-Min Sun}
\affiliation{School of Physics and Astronomy, Sun Yat-sen University, 2 Daxue Road, Tangjia, Zhuhai, Guangdong,  China}

\author[0000-0001-7490-5767]{Zhuang Zhang}
\affiliation{School of Physics and Astronomy, Sun Yat-sen University, 2 Daxue Road, Tangjia, Zhuhai, Guangdong,  China}

\author[0009-0007-1038-4254]{Jun-Yang Liu}
\affiliation{School of Physics and Astronomy, Sun Yat-sen University, 2 Daxue Road, Tangjia, Zhuhai, Guangdong,  China}

\author[0000-0003-2302-0613]{Sheng-Li Qin}
 \affiliation{School of physics and astronomy, Yunnan University, Kunming, Yunnan Province, China}

\author[0000-0002-1086-7922]{Yong Zhang}
\affiliation{School of Physics and Astronomy, Sun Yat-sen University, 2 Daxue Road, Tangjia, Zhuhai, Guangdong,  China}
\affiliation{CSST Science Center for the Guangdong-Hongkong-Macau Greater Bay Area, Sun Yat-Sen University, Guangdong Province, China}



\begin{abstract}

The identification of the 21~$\mu$m feature in some protoplanetary nebulae remains a longstanding puzzle, whose interpretation requires characterization of the molecular gas environments of associated sources.
Here, we present high-resolution interferometric observations from the Northern Extended Millimeter Array toward the 21~$\mu$m protoplanetary nebula IRAS~06530$-$0213. Multiple molecular transitions of HC$_3$N, C$_4$H, and SiC$_2$ are detected and spatially resolved.
We analyze these spectral lines alongside previously reported CO and $^{13}$CO emission toward this source. 
Differing from the extended barrel-like molecular structure traced by CO and $^{13}$CO, HC$_3$N and C$_4$H
 are predominantly concentrated in the inner low-latitude equatorial zones. 
Reproducing the observed morphology requires HC$_3$N to possess a more inwardly concentrated abundance distribution compared with CO and $^{13}$CO, implying that the inner equatorial regions provide favorable conditions for carbon-chain chemistry.

\end{abstract}



\section{Introduction} \label{sec:intro}

Stars with initial masses of 0.8--8 M$_{\odot}$ evolve into the Asymptotic Giant Branch (AGB) phase during their late evolutionary stages. During the AGB, intense dusty stellar winds expel the stellar outer layers at 
mass-loss rates of $\sim10^{-8}$--$10^{-4}$ M$_\odot$ yr$^{-1}$, building 
up a circumstellar envelope around the central star \citep[e.g.,][]{1993ApJ...413..641V,1995A&A...299..755B,2019NatAs...3..408D}. It is widely recognized that AGB envelopes, particularly carbon-rich envelopes, act as active environments for synthesizing diverse molecules.
\citep[e.g.,][]{
2017A&A...601A...4A,2024A&A...684A...4U}.
As the central star’s temperature rises, the circumstellar envelope becomes progressively ionized, and the system transitions into a  planetary nebula (PN). 
Bridging the two evolutionary stages is the short-lived protoplanetary nebula (PPN) phase, with a duration of $\sim10^{3}$ yr, during which the envelope deviates significantly from 
spherical symmetry, displaying bipolar, spiral, or disk-like morphologies \citep{1998AJ....116.1357S,2020Sci...369.1497D}. The physical origin of this diversity remains debated, although recent reviews emphasize the dominant role of binary interactions in sculpting the PPN structure \citep[e.g.,][]{2018Galax...6...99L}.


Aside from variations in morphology, molecular line observations have confirmed the presence of abundant and diverse circumstellar chemistry in PPNs and young PNs \citep[e.g.,][]{1996A&A...315..284H,2001A&A...377..868B,2012ApJS..203...16S}. 
A rich variety of gas-phase molecules have been detected in carbon-rich PPNs such as CRL 2688 and CRL 618 \citep{2007ApJ...661..250P,2008AJ....136.2350P,2013ApJ...773...71Z,2022ApJS..259...56Q}. There is observational evidence for the presence of aromatic compounds, aliphatic species, and fullerenes throughout the PPN phase
\citep{2001ApJ...554L..87K,2011ApJ...730..126Z}. The detection of the 21\,$\mu$m feature further expands the inventory of molecules uniquely identified in PPNs. It remains unclear whether these complex species and the carrier of the 21\,$\mu$m feature form during the AGB phase and are only excited upon transition to the post-AGB phase \citep[e.g.,][]{2016ApJ...825...68M}, or are produced in situ within the PPN stage, a possibility requiring extremely short chemical timescales.
The so-called 21\,$\mu$m feature is an emission feature uniquely detected in carbon-rich PPNs, actually peaking near 20.1\,$\mu$m.
 First identified by  \citet{1989ApJ...345L..51K},
this feature had been detected in only 31 objects \citep[see][for a recent review]{2020Ap&SS.365...88V}.
Despite the proposal of over a dozen candidate carriers ranging from individual carbon- or silicon-bearing molecules to polycyclic aromatic hydrocarbon (PAH) complexes, the 21\,$\mu$m feature \textbf{carrier }remains unidentified \citep[see also][]{2015ApJ...802...39M,2016ApJ...825...68M,2020Ap&SS.365...88V}.
Its exclusive occurrence in carbon-rich PPNs points to a strong link with carbon-chain chemistry, implying that investigations of carbon-bearing molecules in 21\,$\mu$m sources could yield critical insights into identifying this enigmatic feature \citep{2020ApJ...898..151Z,2024AJ....167...91Q}.

A recent radio interferometric study of the 21\,$\mu$m source IRAS 23304$+$6147 using the Northern Extended Millimeter Array (NOEMA)  
reveals that its circumstellar envelope is rich in carbon-chain molecules and silicon-bearing species \citep{2025A&A...696A.102S}.
These molecules trace an equatorial density enhancement (EDE), with longer and more complex carbon chains preferentially located in the outer regions of the EDE. Silicon-bearing species are distributed even farther from the central star than carbon chains, a spatial trend that suggests efficient dust formation within the EDE. For another 21\,$\mu$m source, IRAS~06530$-$0213, a study by \citet{2025AJ....170..231S} indicates that this object entered the protoplanetary nebula (PPN) phase approximately 4500 years ago. 
In CO maps of this source, the outer regions exhibit a spherically symmetric halo inherited from the AGB epoch, whereas the inner regions display a barrel-like  structure, an unambiguous signature of the onset of the PPN phase.
So far, only five 21\,$\mu$m sources have been spatially resolved at high angular resolution with radio interferometric arrays \citep{1999ApJ...524L.125S,2009ApJ...692..402N,2012ApJ...759...61N,2025A&A...696A.102S,2025AJ....170..231S}. All these sources 
exhibit an EDE in their central regions, a morphological structure  commonly seen among PPNs. EDEs modify the local radiation field and govern the excitation and survival of internal molecules. 

As a follow-up to \citet{2025AJ....170..231S}, this paper investigates the molecular species and their spatial structure
within IRAS~06530$-$0213. Observations are briefly described in Sect.\ref{sec:obser}. In Section~\ref{sec:results}, we present imaging results for the detected molecular species.
In Section~\ref{sec:discuss}, we discuss molecular distributions, radiative transfer constraints on excitation and abundance, and the potential link between the inner EDE and the observed infrared emission features.
Our conclusions are summarized in Section~\ref{sec:concl}.

\section{Observations and data reduction} \label{sec:obser}

IRAS~06530$-$0213 was observed with NOEMA in the 10D (project code W20BI) and 12A (project code W21BI) configurations in October 2021 and March 2022, respectively. 
The 10D and 12A datasets differ in frequency coverage and projected baseline range (24--176 m and 32--920 m, respectively), and thus have different angular resolutions. For molecular lines detected in both datasets, we combined the two to construct molecular maps. Prior to merging the visibilities, we removed edge channels, extracted the common frequency ranges (lower sideband, LSB: 214--220.5 GHz; upper sideband, USB: 229.5--236 GHz), and aligned the channel grids.  
Visibilities were calibrated using the \texttt{GILDAS/CLIC} software package\footnote{\url{http://www.iram.fr/IRAMFR/GILDAS}}, while imaging was performed with \texttt{GILDAS/MAPPING} and \texttt{GILDAS/IMAGER}\footnote{\url{https://imager.oasu.u-bordeaux.fr}}. The phase center is defined at the J2000 equatorial coordinates $\left(\alpha_{\rm J2000},\delta_{\rm J2000}\right) = \left(06^{\rm h}55^{\rm m}31\fs820,\, -02^\circ17'28\farcs300\right)$. The synthesized beams of the 10D, 12A, and combined spectral-line cubes are $1.91'' \times 1.29''$, $0.67'' \times 0.3''$, and $0.72'' \times 0.41''$, respectively. The spectral resolution is approximately 2 MHz ($2.7\,\rm km\,s^{-1}$) at 1.3 mm band. The root-mean-square (rms) noise levels of the observed transitions are listed in Table~\ref{line}. We refer the reader to \citet{2025AJ....170..231S} for full details of the calibration, imaging, weighting, and $uv$-coverage procedures.

\section{Results} \label{sec:results}

\begin{figure*}
\centering
\includegraphics[width=0.9\linewidth]{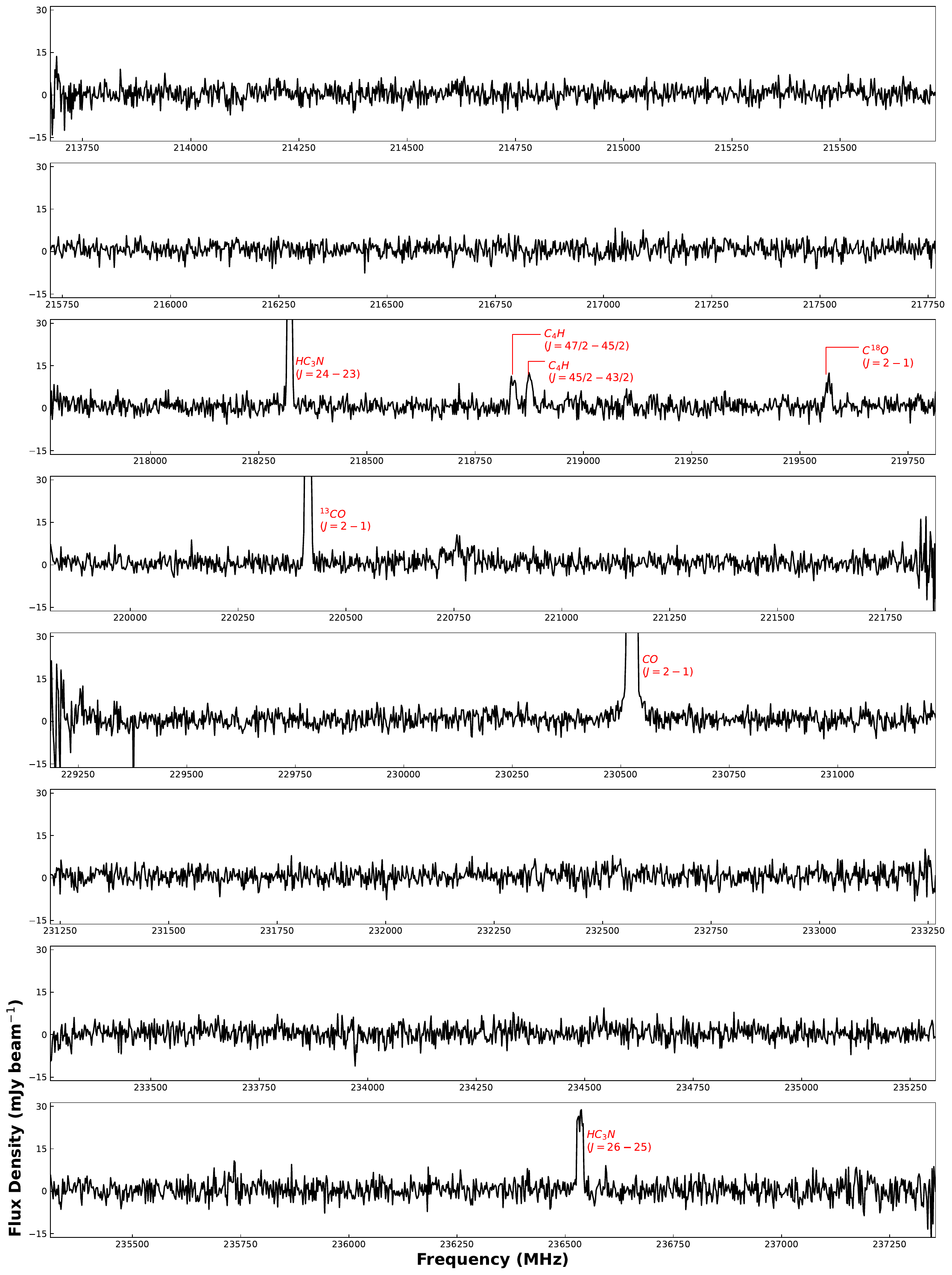}
\caption{NOEMA 10D configuration spectra of  IRAS~06530-0213.}
\label{spec_10d}
\end{figure*}

\begin{figure*} 
\centering
\includegraphics[width=0.9\linewidth]{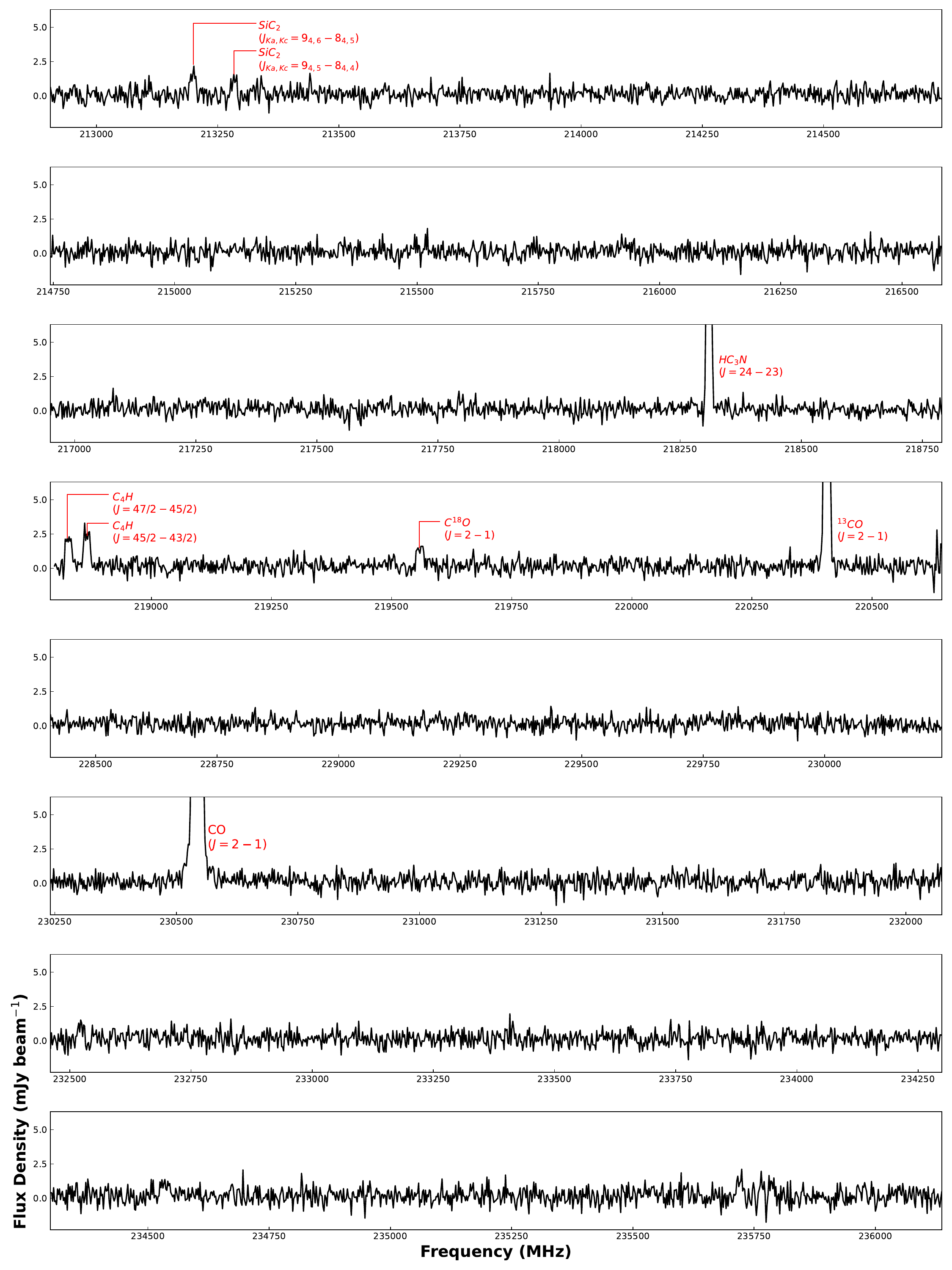}
\caption{NOEMA 12A configuration spectra of  IRAS~06530-0213.}
\label{spec_12a}
\end{figure*}

\begin{figure*} 
\centering
\includegraphics[width=1\linewidth]{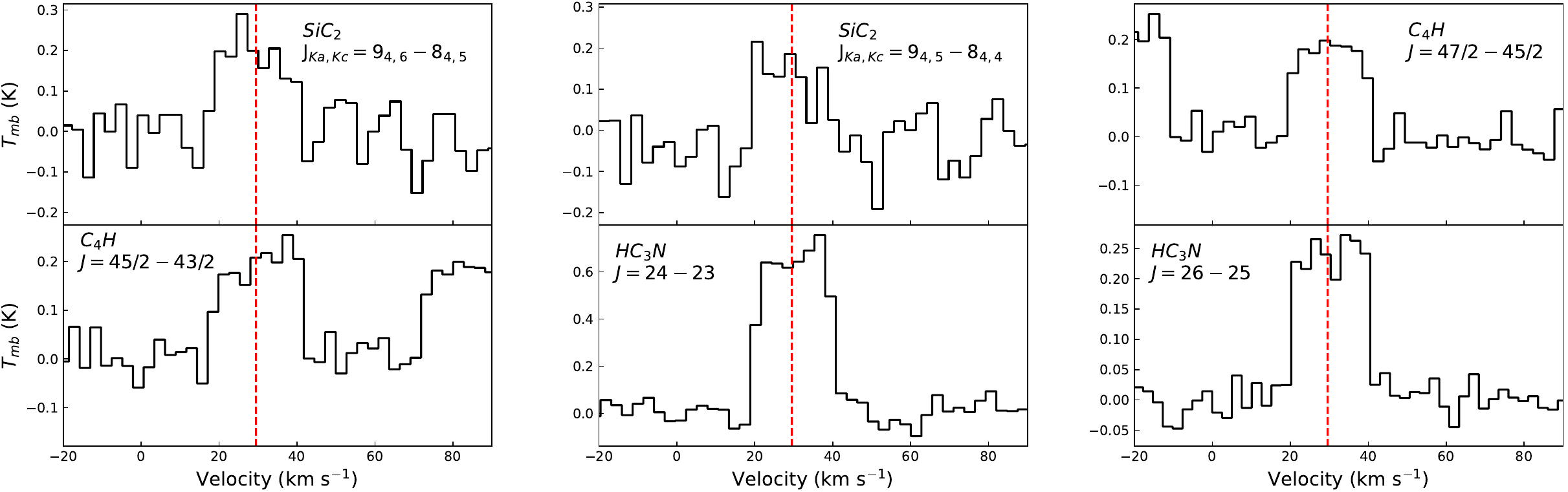}
\caption{Profiles of the detected molecular lines in IRAS~06530$-$0213 (excluding CO and its isotopologues). The systemic velocity of $29.5\,\rm km\,s^{-1}$ is indicated by  vertical dashed lines.}
\label{spec_detail}
\end{figure*}

Including the CO, $^{13}$CO, and C$^{18}$O lines previously reported in \citet{2025AJ....170..231S}, the NOEMA observations reveal nine individual emission features from six molecules and isotopologues.
The majority of these detected species are carbon-bearing molecules.
By averaging the flux density over a $2\arcsec\times2\arcsec$ region at the nebular center, we derived the observed spectra for the 10D and 12A configurations (Figures~\ref{spec_10d} and \ref{spec_12a}), respectively.  All measurements of the spectral lines are summarized in Table~\ref{line}, where $I_{\rm peak}$ and $\int I \, dv$ denote the peak and integrated intensities, respectively. The line profiles are displayed in Figure~\ref{spec_detail}\footnote{The profiles of the $\rm CO$ and $\rm ^{13}CO$ lines have been reported by \citet{2025AJ....170..231S} and are therefore omitted herein.}. In \citet{2025AJ....170..231S}, we derived a systemic velocity of $33\,\mathrm{km\,s^{-1}}$ using the $^{12}$CO $J=2\!-\!1$ line, consistent with the earlier single-dish measurement from \citet{1994A&AS..103..301H}. By contrast, the HC$_3$N and C$_4$H lines analyzed here both exhibit line centroids at $29.5\pm0.2\,\mathrm{km\,s^{-1}}$.
This offset likely arises from velocity gradients induced by expansion motions and the optically thick characteristic of CO lines.
In an expanding envelope where temperature decreases with radius,  CO molecules in the outer layers on the near side preferentially absorb blueshifted emission, which shifts the observed CO peak toward slightly higher velocities. 
Accordingly, we adopt $V_{\rm sys}=29.5\,\mathrm{km\,s^{-1}}$ for all subsequent analyses, with an estimated uncertainty of $0.2\,\mathrm{km\,s^{-1}}$.

In \citet{2025AJ....170..231S}, we presented the spectral and imaging results of the continuum, as well as the $\rm CO$ and $\rm ^{13}CO$ molecules. Specifically, $\rm CO$ exhibits a halo structure with a diameter of approximately $10\arcsec$, surrounding a central barrel-like structure with an outer diameter of ~$2\arcsec$. For $\rm ^{13}CO$, the traced structure in the central region is consistent with that of $\rm CO$, but no halo component is detected. In this work, we report the observational results of other molecules detected in IRAS~06530$-$0213. 


\subsection{HC$_3$N}

\begin{figure*}
    \centering
    \begin{minipage}{0.85\linewidth}
        \centering
        \includegraphics[width=\linewidth]{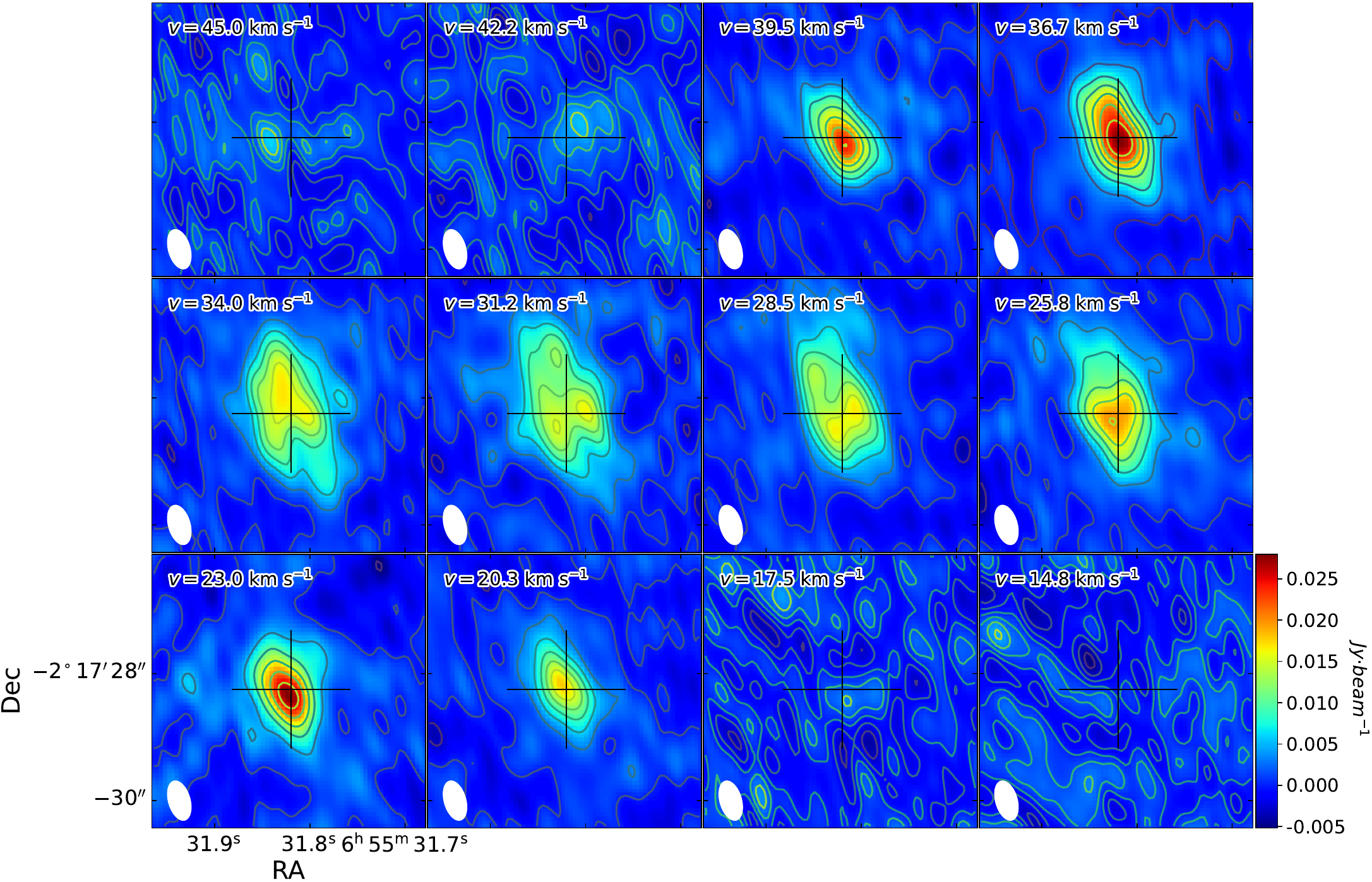}
    \end{minipage}
    \hfill
    \begin{minipage}{0.85\linewidth}
        \centering
        \includegraphics[width=\linewidth]{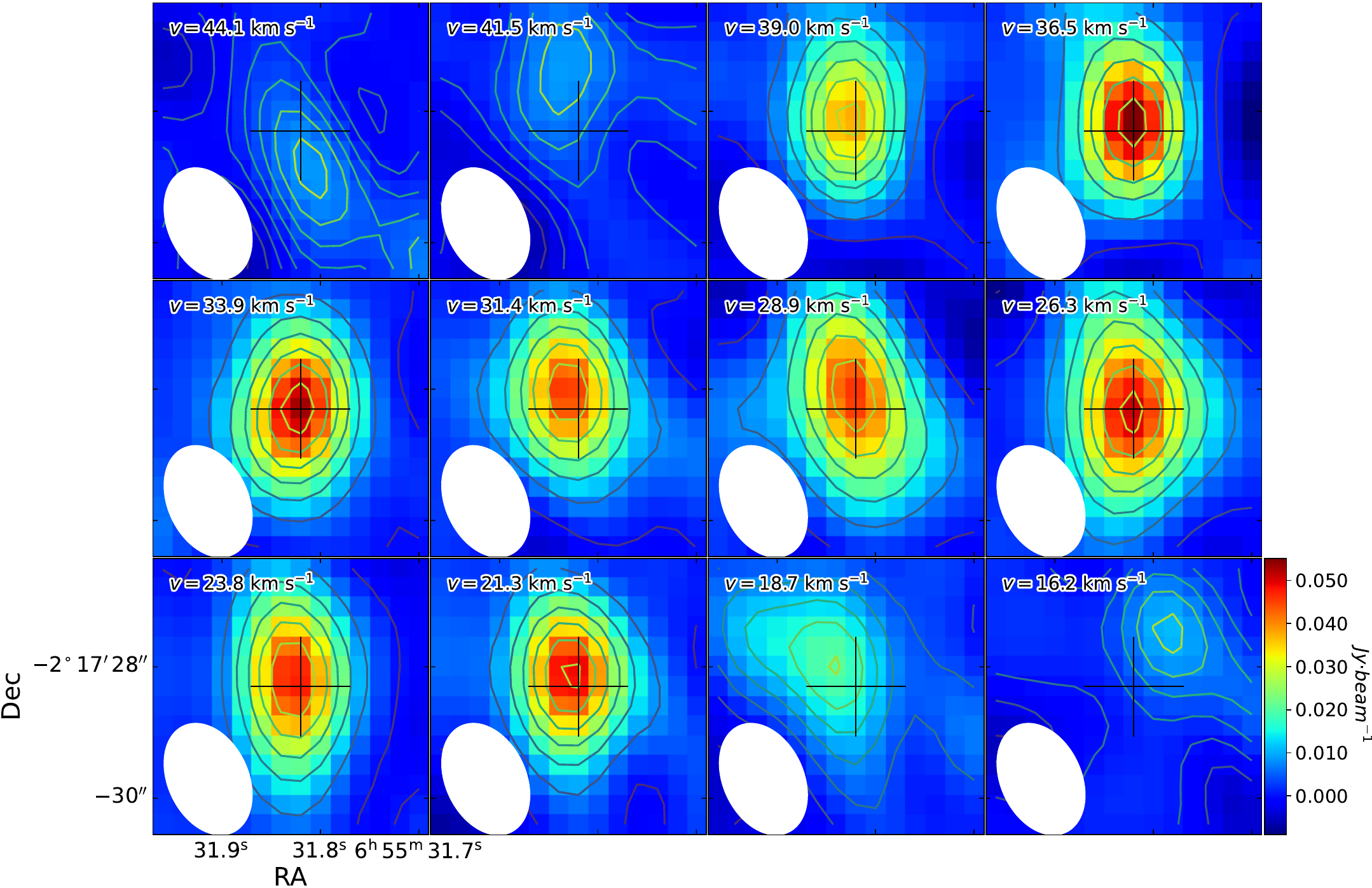}
    \end{minipage}
    \caption{
    Channel maps of HC$_3$N. Contour levels are individually defined for each channel by ten equally spaced lines between the channel's intensity minimum and maximum. Crosses mark the phase center position, and LSR velocities are labeled in each panel. The object's systemic velocity is $29.5\,\rm{km~s^{-1}}$. The white ellipse in the bottom-left corner denotes the synthesized beam. The upper panels are derived from combined data and shows the emission of the $J=24\!-\!23$  transition with a synthesized beam of $0.72\arcsec\times 0.41\arcsec$, while the lower panels come from observations with the 10D configuration and displays the emission of the $J=26\!-\!25$  transition with a synthesized beam of $1.91\arcsec\times 1.29\arcsec$.
    }
    \label{hc3n_channel}
\end{figure*}

Figure~\ref{hc3n_channel} shows the continuum-subtracted channel maps of $\rm HC_3N$ emission for the $J=24\!-\!23$ and $J=26\!-\!25$ transitions. As shown in the upper panels,
which correspond to higher-spatial-resolution data, 
 \(\rm HC_3N\) (\(J=24\!-\!23\)) traces a faint torus-like structure across the velocity range of 25.8--34.0\,km~s\(^{-1}\). This structure has an outer diameter of approximately \(1\arcsec\) and extends along the southeast–northwest axis.
This line preferentially traces the inner, low-latitude regions of the barrel-like structure delineated by CO and $^{13}$CO
\citep{2025AJ....170..231S}.
For the $\rm HC_3N$ ($J=26\!-\!25$) transition, observations were performed exclusively with the compact 10D array configuration. The correspondingly low spatial resolution prevents robust measurements of additional spatial structural properties for this line.

\begin{figure*} 
\centering
\includegraphics[width=\linewidth]{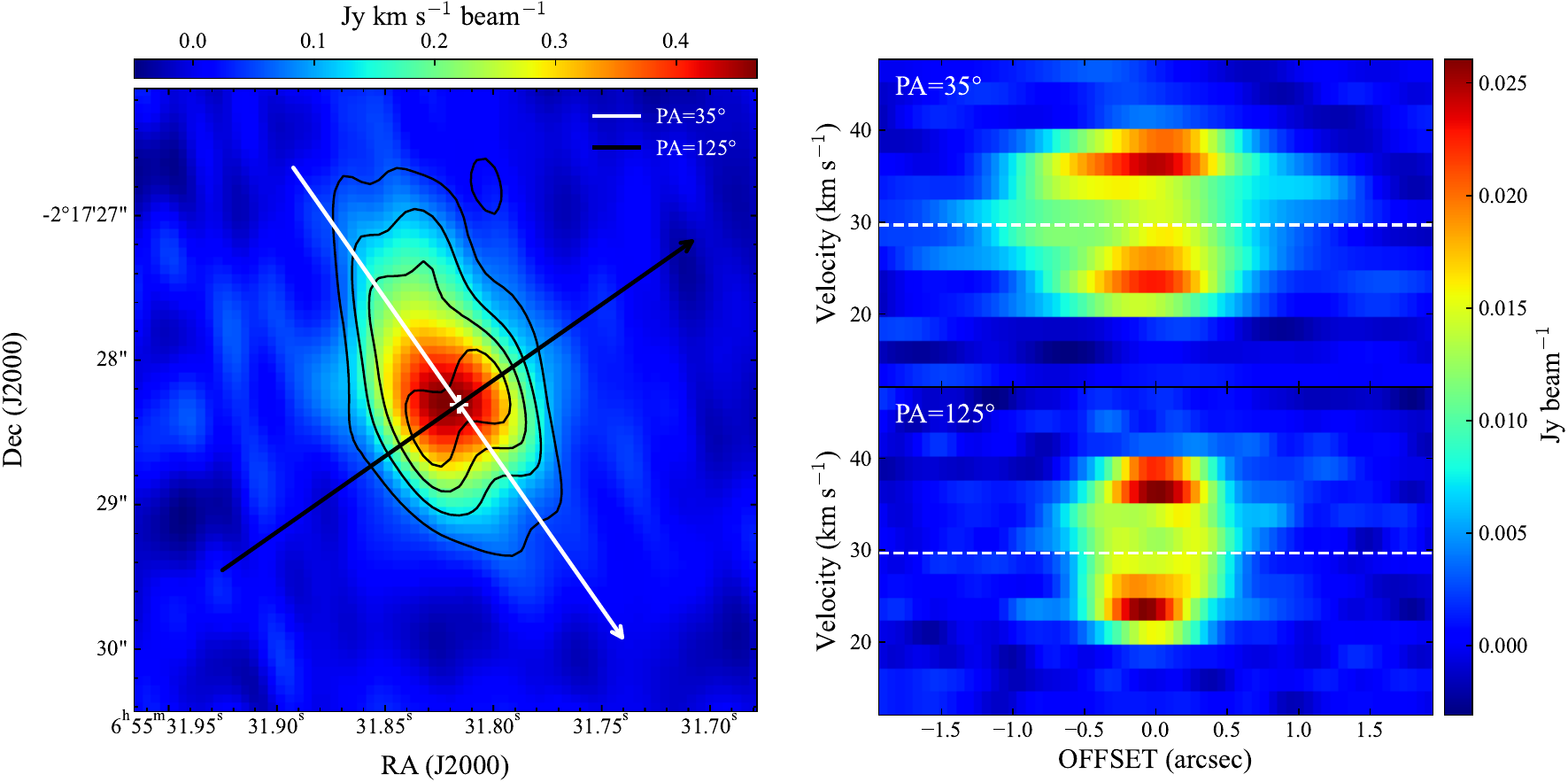}
\caption{{\it Left}: Moment-zero map of HC$_3$N ($J=24\!-\!23$) integrated over the full velocity range(color scale), with moment-zero contours overlaid from channels showing torus-like morphology
 (25.8--34.0~km~s$^{-1}$). Contour levels are defined as  25\%, 45\%, 65\%, and 85\% of the peak intensity. White and black arrows mark the directions for extracting PV diagrams
 at  position  angles  of PA$=35^\circ$ and $125^\circ$, 
 which follow the symmetry axis and equatorial direction of the barrel-like structure, respectively.
 {\it Right}: PV diagrams extracted along the two directions. Dashed lines mark  $V_{\mathrm{LSR}} = 29.5\ \mathrm{km\ s^{-1}}$. Offsets are measured from the arrow centers, with negative values toward the tails and positive values toward the arrowheads.}
\label{hc3n_pv}
\end{figure*}

Figure~\ref{hc3n_pv} presents the position–velocity (PV) diagrams of the HC$_3$N ($J=24\!-\!23$) line. The overall emission is dominated by the torus-like structure, with only faint contributions from the extended lobes. Along the nebular equatorial direction, the PV diagram exhibits an elliptical ring morphology, while the PV cut along the symmetry axis displays two distinct bright emission features on the blue and red sides of the systemic velocity, respectively. This characteristic signature is indicative of an expanding torus dominating the HC$_3$N emission. Fainter, elongated lobe structures are also visible along the axial direction, with line-of-sight velocities lower than those of the central torus.
The centrally symmetric distribution of emission in the PV diagram suggests that the central torus is viewed nearly edge-on. Prior CO  observations presented in \citet{2025AJ....170..231S} yield an inclination angle of \(20^\circ\), with the southwest lobe oriented toward the observer for this source system.
This small inclination is barely discernible in the HC$_3$N PV diagram, as HC$_3$N traces predominantly the innermost gaseous regions of the structure  in comparison with CO. 
Adopting the same inclination angle, we estimate the deprojected expansion velocity of the torus to be $\sim10.6$~km~s$^{-1}$.

\subsection{C$_4$H}

\begin{figure*}
    \centering
    \begin{minipage}{0.8\linewidth}
        \centering
        \includegraphics[width=\linewidth]{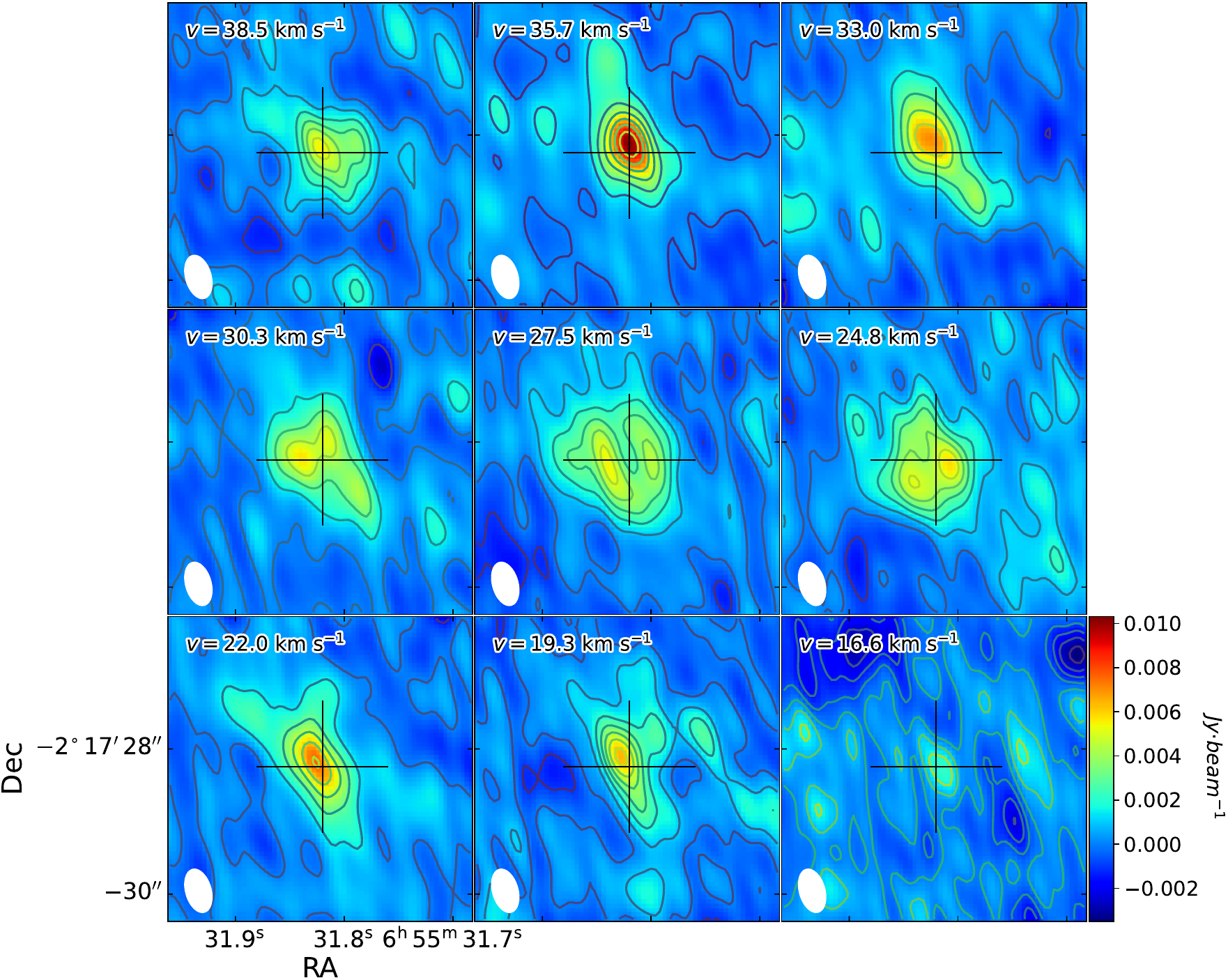}
    \end{minipage}
    \hfill
    \begin{minipage}{0.8\linewidth}
        \centering
        \includegraphics[width=\linewidth]{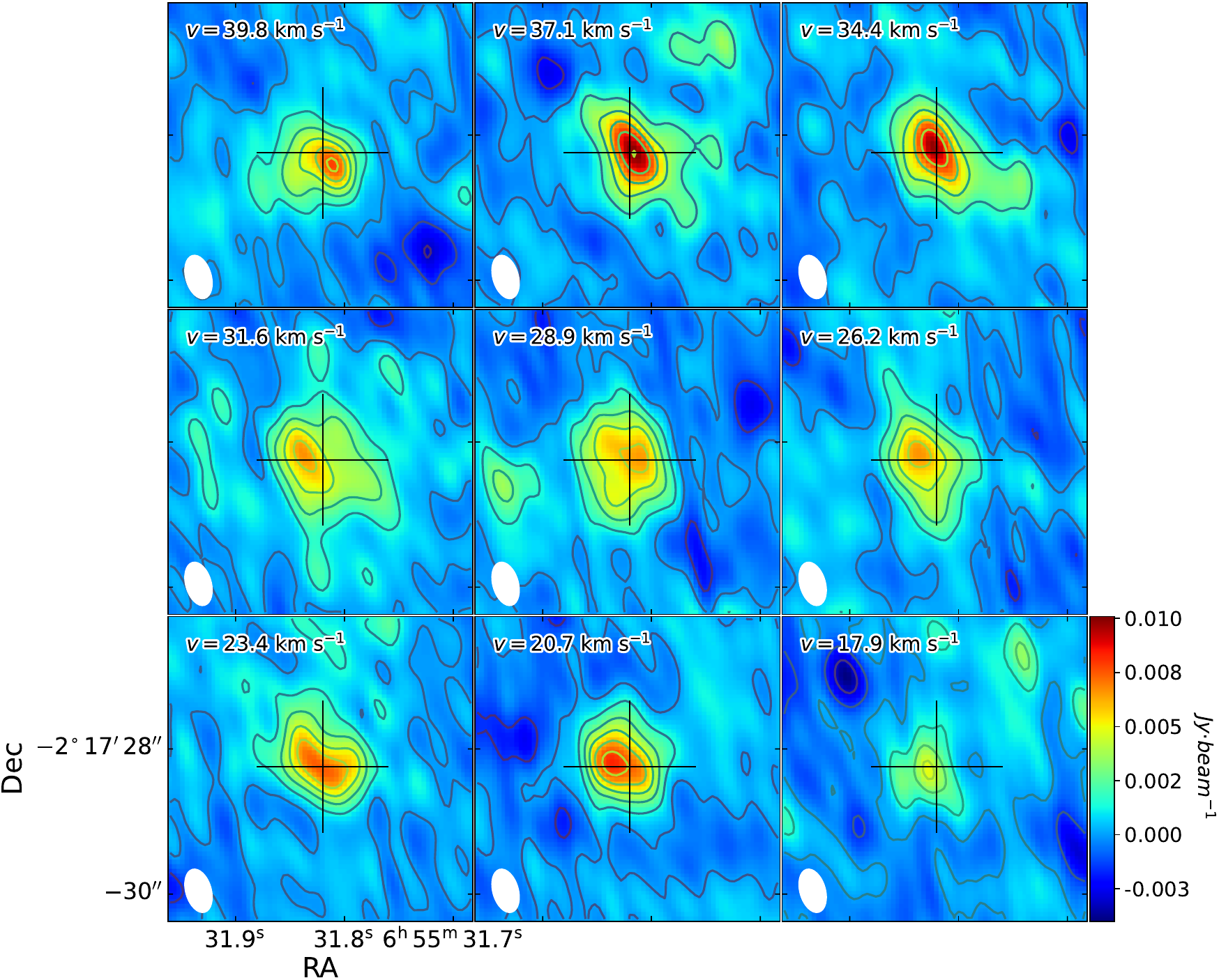}
    \end{minipage}
    \caption{Same as the upper panel of Figure~\ref{hc3n_channel}, but showing results for C$_4$H $J=47/2$--$45/2$ (upper) and $J=45/2$--$43/2$ (lower).}
    \label{c4h_channel}
\end{figure*}

\begin{figure*} 
\centering
\includegraphics[width=\linewidth]{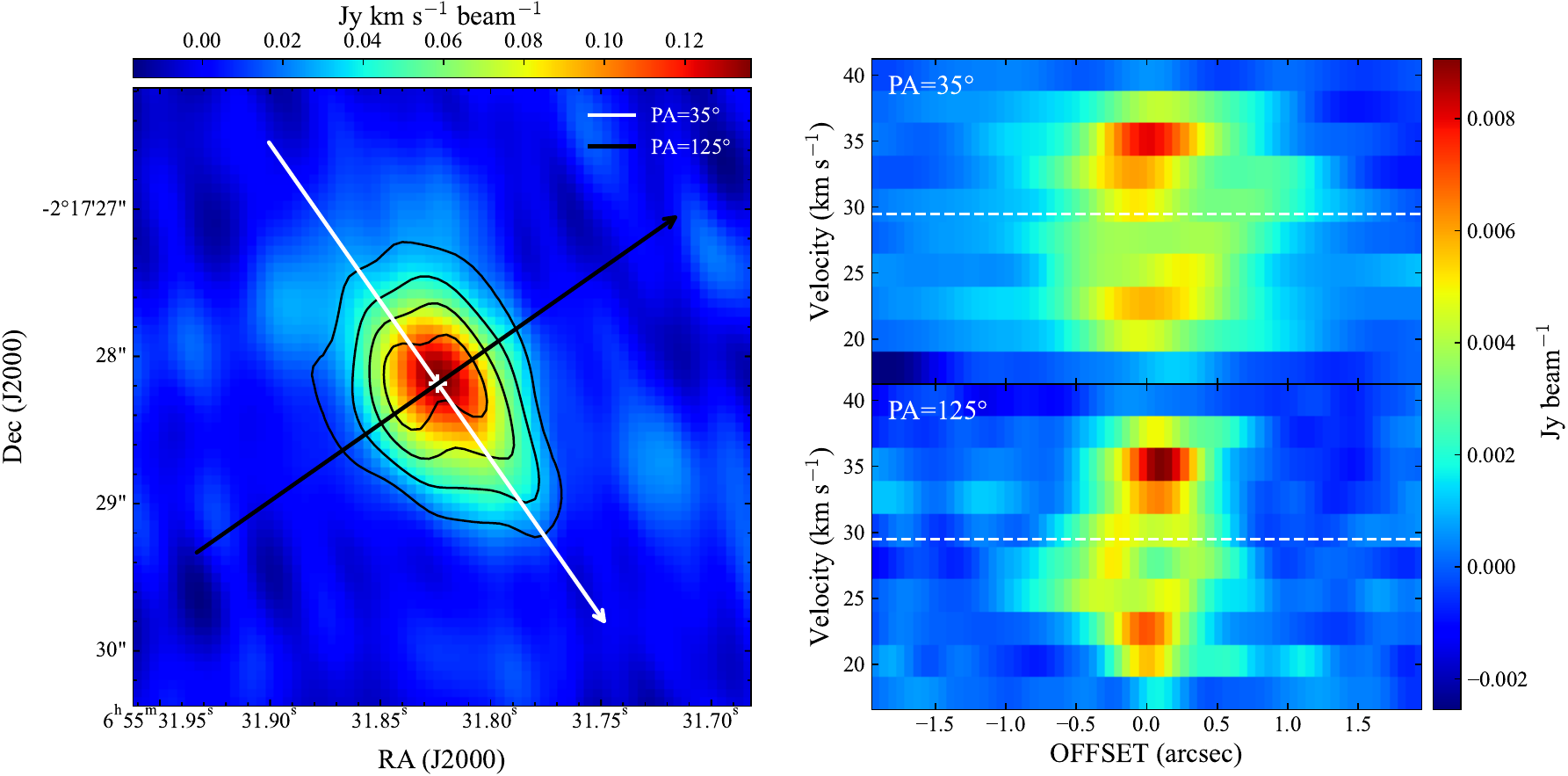}
\caption{ Same as the Figure~\ref{hc3n_pv}, but for C$_4$H ($J=47/2$--$45/2$). }
\label{c4h_pv}
\end{figure*}

Figure~\ref{c4h_channel} presents the continuum-subtracted channel maps of C$_4$H emission for the $J=47/2-45/2$ and $J=45/2-43/2$ transitions. Within the velocity interval $24$--$31\,\mathrm{km\,s^{-1}}$, emission from the $\rm C_4H$ $J=47/2-45/2$ transition delineates a flat torus-like structure extending from southeast to northwest. It predominantly traces the inner, low-latitude zones of the barrel structure seen in $\rm CO$ and $\rm ^{13}CO$. The corresponding PV diagram is displayed in Figure~\ref{c4h_pv}. Kinematically, $\rm C_4H$ shows overall features similar to $\rm HC_3N$, but the emission is fainter and concentrated in the innermost material.
This torus-like feature is not clearly resolved in the $J=45/2-43/2$ transition, mainly owing to limited spatial resolution and observational sensitivity.


\subsection{SiC$_2$}

\begin{figure*} 
\centering
\includegraphics[width=0.9\linewidth]{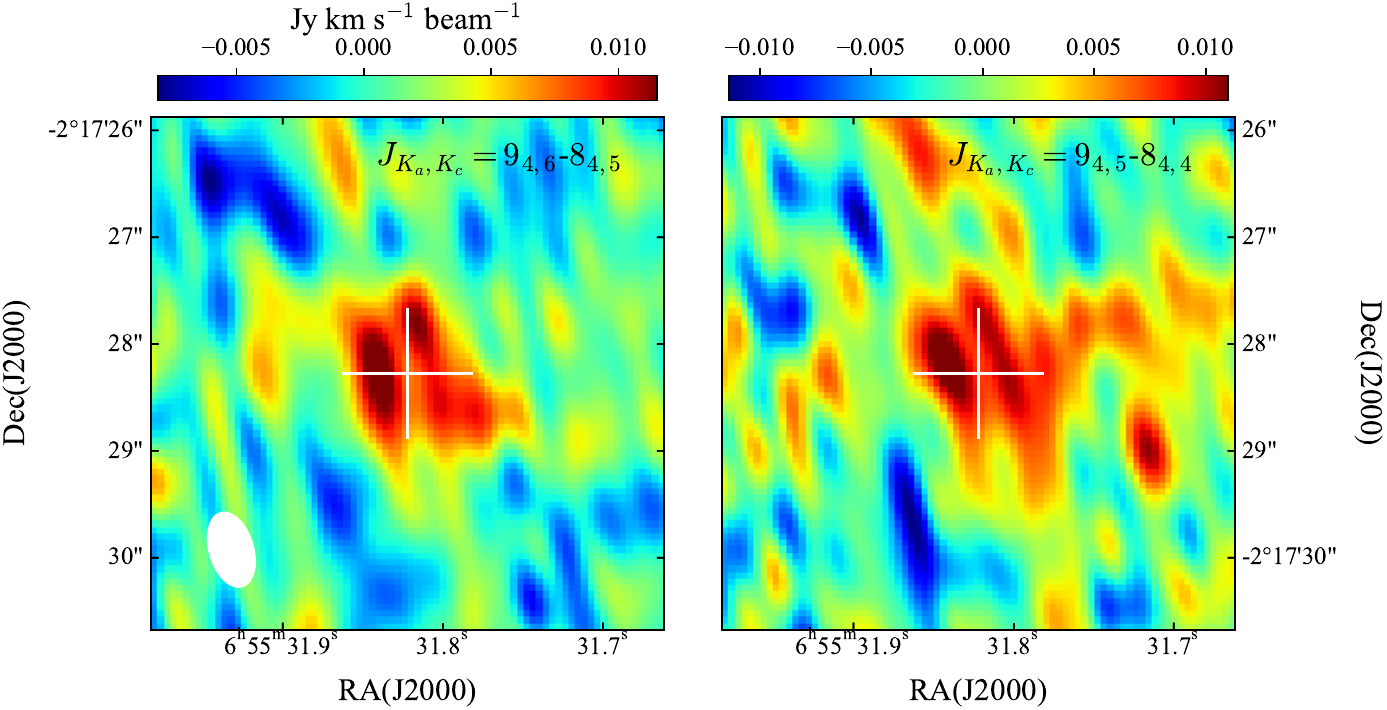}
\caption{Moment-zero maps of the SiC$_2$  
$J_{K_a,K_c}=9_{4,6}$--$8_{4,5}$ (left panel)
and $J_{K_a,K_c}=9_{4,5}$--$8_{4,4}$ (right panel) transitions.
Crosses denote the phase center. The synthesized beam, shown in the bottom-left corner, has a size of $0.72\arcsec\times 0.41\arcsec$.}
\label{sic2}
\end{figure*}

\begin{figure*} 
\centering
\includegraphics[width=1\linewidth]{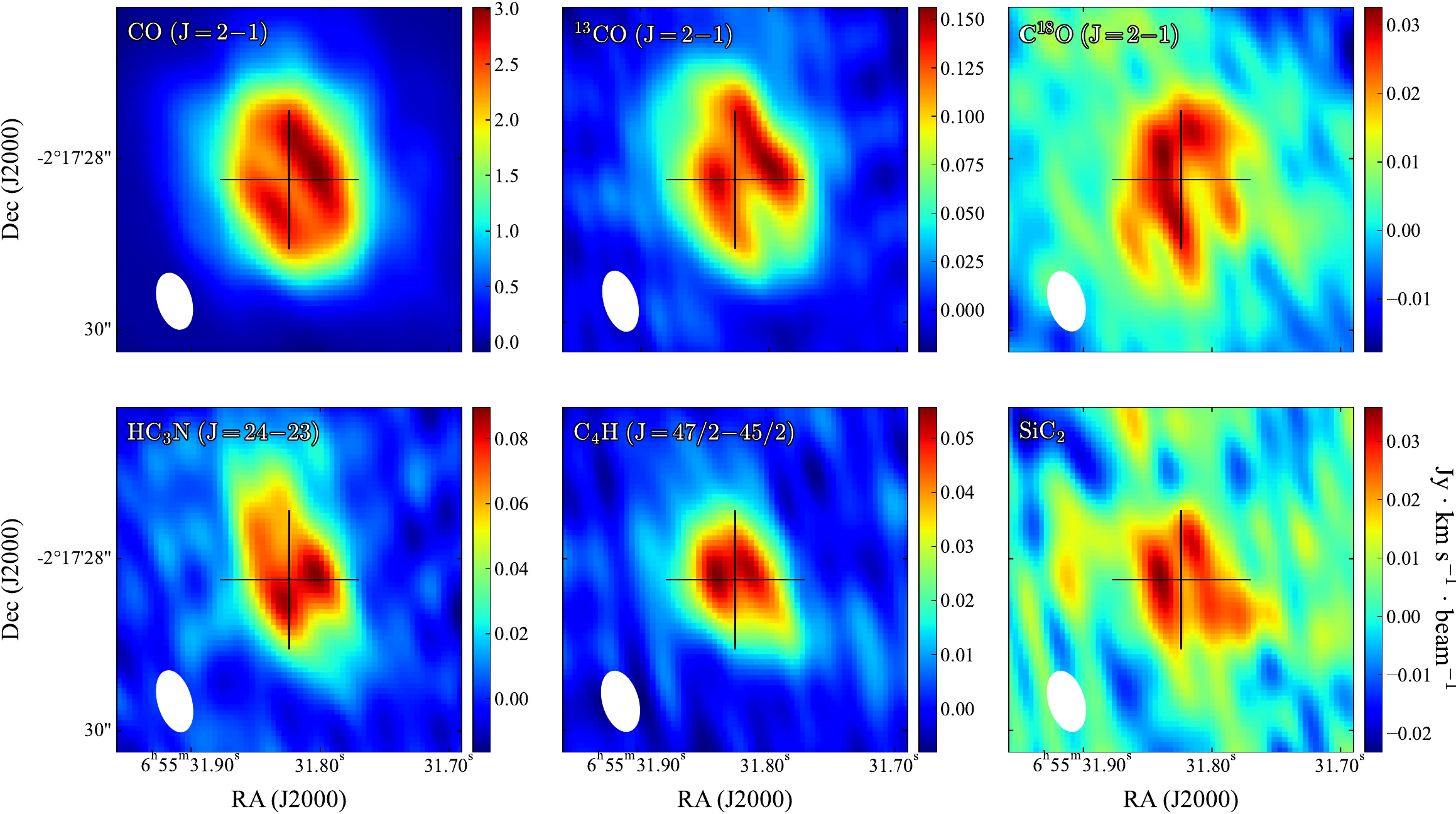}
\caption{Comparison of moment-zero maps for detected molecular lines in IRAS~06530$-$0213. Crosses mark the phase center, and the white ellipse in the bottom-left corner denotes the synthesized beam ($0.72\arcsec\times0.41\arcsec$).
The map of SiC$_2$ is produced by stacking the two maps shown in
Figure~\ref{sic2}.}
\label{molecules}
\end{figure*}

We clearly detected two faint $\rm SiC_2$ transitions. Owing to the low signal-to-noise ratio (SNR), individual channel maps contain little usable information. The moment-zero maps for the two transitions are presented in Figure~\ref{sic2}. 
Since the two lines share nearly identical rest frequencies and exhibit consistent excitation properties, we resampled all datasets to a common velocity grid and carried out uv-plane visibility combination to construct a co-added \(\rm SiC_2\) moment map, as illustrated in Figure~\ref{molecules}.

The stacked $\rm SiC_2$ emission broadly matches the barrel-like molecular structure traced by $\rm CO$ and $\rm ^{13}CO$. However, due to the low SNR of the $\rm SiC_2$ data, its morphology cannot be characterized with the same detail as the $\rm CO$ and $\rm ^{13}CO$ structures. The detected $\rm SiC_2$ emission is predominantly concentrated in the equatorial regions of the barrel structure and is elongated along the northeast–southwest direction, rather than tracing the full extent of the barrel walls.




\subsection{Molecular Abundances}
Under the assumptions of optically thin emission and local thermodynamic equilibrium (LTE), we derive the excitation temperature ($T_{\rm ex}$) and total column density ($N_{\rm tot}$) for HC$_3$N. The population of the upper level $N_u$ is given by 
\begin{equation}
N_u=\frac{8\pi k \nu^2}{h c^3 A_{ul}}\int I\,dv,
\label{eq:NuW}
\end{equation}
where $\nu$ denotes the rest frequency, $A_{ul}$ represents the Einstein $A$ coefficient, and $\int I\,dv$ corresponds to the velocity-integrated intensity.
$T_{\rm ex}$ and $N_{\rm tot}$ are derived via  the following relation: 
\begin{equation}
\frac{N_u}{g_u}=\frac{N_{\rm tot}}{Q(T_{\rm ex})}\exp\!\left(-\frac{E_u}{kT_{\rm ex}}\right),
\label{eq:LTEpop}
\end{equation}
where $g_u$ is the statistical weight,  $E_u$ is the upper-level energy, and $Q(T_{\rm ex})$ is the partition function. 
The two $\rm HC_3N$ lines were observed with distinct array configurations, yielding different synthesized beams. To enable reliable direct comparison, the two datasets were processed individually from their respective uv tables while adopting identical imaging parameters. These parameters were set to match those used for the $J=26$–$25$ data cube from the 10D configuration observations, and all final images were restored with a uniform synthesized beam. This calibration and imaging strategy ensures that both transitions trace consistent spatial scales before spectral extraction and line integration.
We obtain $T_{\rm ex}=24\pm2$~K and  $N_{\rm tot}=(4.5\pm1.3)\times10^{14}~\mathrm{cm^{-2}}$.

For other detected molecular species,  excitation temperatures  cannot be derived. We therefore adopt a fixed value of $T_{\rm ex}=24$~K and calculate the total column density $N_{\rm tot}$ using Equation~(\ref{eq:NuW}).
The resulting column densities are $(3.7\pm0.8)\times10^{15}~\mathrm{cm^{-2}}$ for $^{13}$CO, $(3.9\pm1.9)\times10^{12}~\mathrm{cm^{-2}}$ for SiC$_2$, and $(1.3\pm0.1)\times10^{15}~\mathrm{cm^{-2}}$ for C$_4$H.
The SiS and CH\(_3\)CN (\(J=12\)–11) transitions at 217.8 and 220.7\,GHz fall within our observed frequency range yet are too weak to be detected. Using the measured rms noise level, we derive upper limits of \(N(\text{SiS}) < 2.4\times10^{13}~\mathrm{cm^{-2}}\) and \(N(\text{CH}_3\text{CN}) < 1.0\times10^{13}~\mathrm{cm^{-2}}\) for the column densities.

We computed the fractional abundance of each molecule, defined as
$X\equiv N_{\rm tot}/N({\rm H_2})$. 
 The column density of molecular hydrogen
 was estimated from the $^{13}$CO ($J=2$--1) line under the LTE and optically thin assumptions. The derivation follows the relation
 \begin{equation}
N({\rm H_2})=\frac{N_{\rm tot}(^{13}{\rm CO})\,R}{X({\rm CO})},
\label{eq:NH2from13CO}
\end{equation}
where $R$ is the carbon isotopic ratio ${}^{12}{\rm C}/{}^{13}{\rm C}$.
Adopting a representative carbon-star isotopic ratio of 
 $R=30$ \citep{2014A&A...566A.145R},
 and $X({\rm CO})=8\times10^{-4}$ \citep[a representative value for 
carbon-rich AGB envelopes,][]{2022MNRAS.510.1204V}, we derive
   $N({\rm H_2})=(1.39\pm0.30)\times10^{20}~{\rm cm^{-2}}$.
Table~\ref{tab:tex_ncol} lists the derived fractional abundances, which are compared with published results for other C-rich circumstellar envelopes possessing or lacking the 21\,\(\mu\)m feature.
For the non-detected species, the molecular abundances \(X(\text{SiS})\) and \(X(\text{CH}_3\text{CN})\) are constrained to be less than \(1.7\times10^{-7}\) and \(7.2\times10^{-8}\), respectively.
 For reference, we further incorporate IRC+10216, a carbon-rich AGB envelope widely used as a benchmark for circumstellar chemistry research.
 We find no systematic differences in molecular abundances across these sources. However, the 21\,\(\mu\)m  sources  tend to show a higher HC$_3$N/SiC$_2$ abundance ratio. 
 We tentatively suggest that these sources may produce more cyanopolyynes, and silicon-bearing molecules are more effectively depleted onto dust grains.

\section{Discussion}
\label{sec:discuss}

The transition from the AGB to the post-AGB phase is marked by dramatic changes in circumstellar chemistry. Molecules synthesized in the cool AGB envelope are either reprocessed or progressively destroyed as the central star becomes hotter and emits intense ultraviolet radiation, while new molecular species can form in the resulting PPN. 
Building on our first detection and spatial mapping of several key molecular species in IRAS 06530$-$0213, we here investigate their spatial distributions and chemical characteristics. 
Our goal is to constrain the molecular architecture of this object and evaluate whether these chemical tracers can shed light on the still poorly understood physical environment associated with the 21\,\(\mu\)m feature.

\subsection{Molecular distributions and radiative transfer constraints}
\label{4.1}

Figure~\ref{molecules} displays moment-zero maps of all detected molecules, produced by integrating emission over the velocity range 24--33~km\,s\(^{-1}\), which corresponds to the signature of the barrel-like structure. All these molecules reside within the same barrel-shaped molecular structure of IRAS~06530$-$0213 but differ noticeably in their spatial concentration. CO and its isotopologues trace the entire nebular structure, whereas HC\(_3\)N and C\(_4\)H peak in the inner equatorial zones. Given its poor SNR, the SiC\(_2\) emission merely follows this general morphological trend.

These maps cannot directly constrain the intrinsic chemical abundance structure. Differences in the spatial distribution of the observed molecules can additionally stem from variable excitation conditions and optical depth variations. We performed simplified one-dimensional radiative transfer simulations using the RATRAN code, a Monte Carlo molecular line radiative transfer solver that calculates molecular excitation and generates synthetic line emission for predefined density, temperature, velocity and abundance distributions \citep{2000A&A...362..697H}. Our goal is not to replicate the complex non-spherical morphology of IRAS~06530$-$0213, but to establish a one-dimensional benchmark to explore how radial profiles are modulated by excitation effects and adopted effective abundance distributions. Although C\(_4\)H is present in this source, it was not included in our RATRAN modelling, owing to the insufficient availability of collisional excitation data required for non-LTE calculations of its detected transitions.

\begin{figure*} 
\centering
\includegraphics[width=0.9\linewidth]{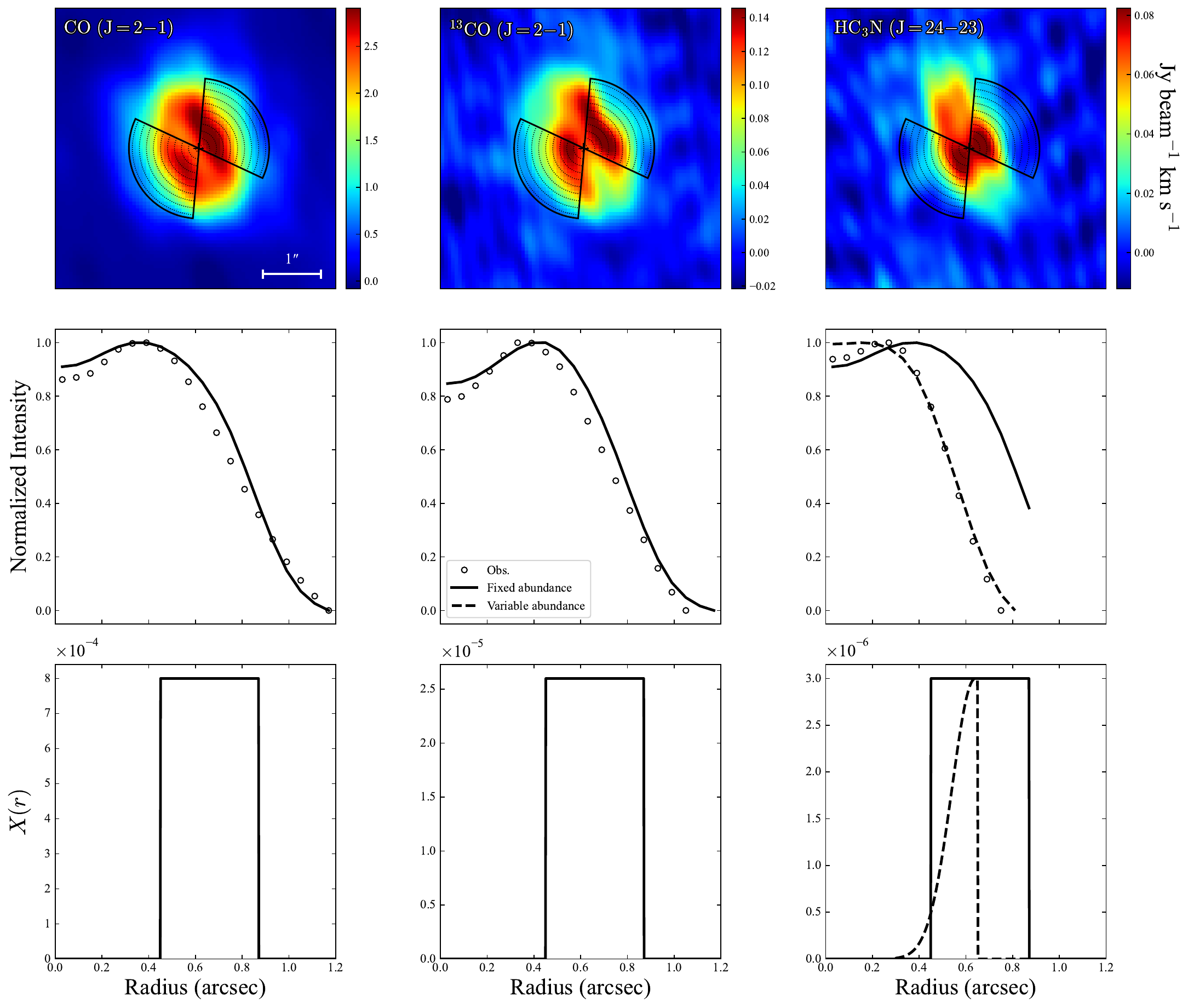}
\caption{
Modeling results from RATRAN calculations compared with observations for CO (left), \(^{13}\)CO (middle), and HC\(_3\)N (right). 
The upper panels show moment-zero maps with the equatorial sector extraction regions overlaid.
 The middle panels compare normalized radial profiles from observations with the corresponding model predictions, where the model profiles have been convolved with the telescope beam.
The lower panels show the radial abundance profiles adopted in our models.}
\label{ratran_profiles}
\end{figure*}

We constructed models for CO, $^{13}$CO and HC$_3$N, adopting a common global physical structure for all three molecular species.
The radial H$_2$ number density follows a power-law profile, 
$n_{\rm H_2}(r) = n_{\rm out}\left(\frac{r}{R_{\rm out}}\right)^{-2}$,
where $n_{\rm out}$ and $R_{\rm out}$ correspond to the H$_2$ number density and radius at the outer boundary, respectively.
We assume a homologous expansion velocity field of the form
$v(r) = v_{\rm max}\left(\frac{r}{R_{\rm out}}\right)$,
while the kinetic temperature of the gas
 is parameterized as
$T(r) = T_{\rm out}\left(\frac{r}{R_{\rm out}}\right)^{-0.4}$,
where $v_{\rm max}$ and $T_{\rm out}$ represent the expansion velocity and gas temperature at the outer boundary, respectively.
For comparative analysis, we extracted radial brightness distributions from the observed moment-zero maps at positions roughly aligned with the equatorial direction  (Figure~\ref{ratran_profiles}).
We averaged the emission across radial bins within two opposite equatorial sectors centered at the source position. These sectors, used consistently for CO, \(^{13}\)CO and HC\(_3\)N, extend to a radius of \(1.2''\) with a radial bin width of \(0.06''\).
 As shown in Figure~\ref{ratran_profiles}, HC\(_3\)N emission is more centrally concentrated than that of CO and \(^{13}\)CO. We determined the radial extent using the outermost radial bins with a SNR \(\geq 3\) from the equatorial sector profiles. Along the equatorial direction, CO and \(^{13}\)CO emission is detectable up to \(\sim1.2''\), while HC\(_3\)N presents a more compact spatial distribution, with its prominent emission limited to within \(\sim0.8''\).

The radial profile of CO emission was adopted to constrain the physical structure for the  calculations. Thanks to its bright emission, high SNR and efficient self-shielding, CO acts as a more robust tracer of the bulk molecular gas than fainter molecular species. We therefore first fitted the normalized CO profile by assuming a constant CO abundance of \(X({\rm CO})=8.0\times10^{-4}\).
We find that our adopted parameters, namely \(n_{\rm H_2,out}=2.0\times10^{5}~{\rm cm^{-3}}\), \(T_{\rm out}=60\) K, and an effective CO-emitting shell spanning from \(R_{\rm in}({\rm CO})=0.45''\) to \(R_{\rm out}({\rm CO})=0.87''\), successfully reproduce the observed CO radial profile.
We note that this one-dimensional model is not designed to yield a unique reconstruction of the source’s intrinsic non-spherical structure. Its primary purpose is to provide a standardized physical benchmark against which we compare the radial profiles of CO, \(^{13}\)CO, and HC\(_3\)N. With the CO-constrained temperature and density structure in hand, we proceeded to examine two scenarios for \(^{13}\)CO and HC\(_3\)N: reproduction via constant abundances, or the inclusion of radial abundance variations.


We adopted empirical abundances of \(X(^{13}{\rm CO})=2.6\times10^{-5}\) and \(X({\rm HC_3N})=3\times10^{-6}\) for the fixed-abundance models. The model results are presented in Figure~\ref{ratran_profiles}. The radial emission profile of \(^{13}\)CO is well reproduced by the model. In contrast, the model fails to match HC\(_3\)N observations, as its predicted profile extends considerably farther outward than the actual spatial distribution of HC\(_3\)N. This indicates that excitation effects or optical depth alone cannot explain the central concentration of HC\(_3\)N, implying that radial variations in molecular abundance are necessary.
Adopting a variable abundance for HC\(_3\)N confined between \(R_{\rm in}=0.30''\) and \(R_{\rm out}=0.65''\) with a peak at \(r=0.64''\), the model can reasonably reproduce the observations. This suggests that chemical differentiation may govern the spatial segregation of molecular species. Relative to CO, HC\(_3\)N is predominantly produced in the inner equatorial region.

\subsection{Possible relevance of the inner EDE to infrared emission features}

The RATRAN experiments suggest that the inner EDE of IRAS~06530$-$0213 may provide physical and chemical conditions favorable for the formation and survival of carbon-chain gas-phase molecules. 
This possibility is broadly consistent with what has been inferred in other carbon-rich evolved objects. 
For example, the carbon-rich PPN CRL~618 hosts detected hydrocarbon species including C\(_4\)H\(_2\), C\(_6\)H\(_2\), and C\(_6\)H\(_6\), which are interpreted as products of photon- and shock-driven chemistry within its dense central region \citep{2001ApJ...546L.123C}. Dedicated chemical models for CRL~618 further demonstrate that high temperatures, high densities, and elevated ionization levels promote efficient ion–molecule reactions, thereby facilitating the formation of benzene and other carbon-rich molecular species \citep{2002ApJ...574L.167W,2003A&A...402..189W}.
Generally, in carbon-rich circumstellar environments, photodissociation of parent molecules such as HCN and C$_2$H$_2$ can produce radicals such as CN and C$_2$H, which subsequently promote carbon-chain and cyanopolyyne chemistry \citep{2017A&A...601A...4A}. 
A natural question is therefore whether the inner EDE
traced by HC\(_3\)N could also regulate the
chemical processing, excitation, or survival of larger carbonaceous compounds and complex organics tied to the infrared emission features.

Although radio observations indicate that the gas-phase molecular inventory in PPNs is less chemically rich than that in AGB envelopes \citep[e.g.][]{2006PNAS..10312274Z,2022EPJWC.26500029A}, aromatic/aliphatic features and fullerenes are nevertheless detected in some PPNs, in contrast to their absence in AGB envelopes. One possible interpretation is that the altered physical conditions of the post-AGB phase, especially the stronger ultraviolet radiation field, may promote the further processing or growth of large, complex organic molecules. An alternative possibility, however, is that some of their carriers are already synthesized during the AGB phase and become observable only in the PPN stage, when a hotter central star provides the ultraviolet photons needed for their excitation \citep{2016ApJ...825...68M}. Although the relative roles of inheritance from the AGB phase and subsequent processing during the PPN stage remain uncertain, 
EDE structures  may provide a favorable high-density environment in which AGB-formed species can survive into the PPN stage. Under the stronger ultraviolet radiation field of the PPN phase, this inherited material may then be further processed, excited, or participate in the formation of more complex molecules. 
EDEs are routinely detected around post-AGB stars exhibiting near-infrared excesses, primarily through observations of CO line emission \citep{2001A&A...377..868B}. The inner EDE regions likely host photodissociation regions (PDRs), where CO is photodissociated to generate free atomic carbon. This chemical pathway promotes the production of carbon-chain molecules and complex organic species, including PAHs \citep{2004ApJ...608L..41C,2014MNRAS.441..364G,ba23}. This established scenario has been widely adopted to interpret PAH formation in the planetary nebula NGC\,6720 \citep{ring1,ring2}.
Our RATRAN modelling simulations support this physical framework. The preferential concentration of HC$_3$N in the inner EDE implies that this region offers favorable conditions for the synthesis of carbonaceous material. However, we cannot fully rule out the alternative scenario that the carriers of infrared emission features are already formed during the prior AGB evolutionary phase.

\begin{figure*} 
\centering
\includegraphics[width=1\linewidth]{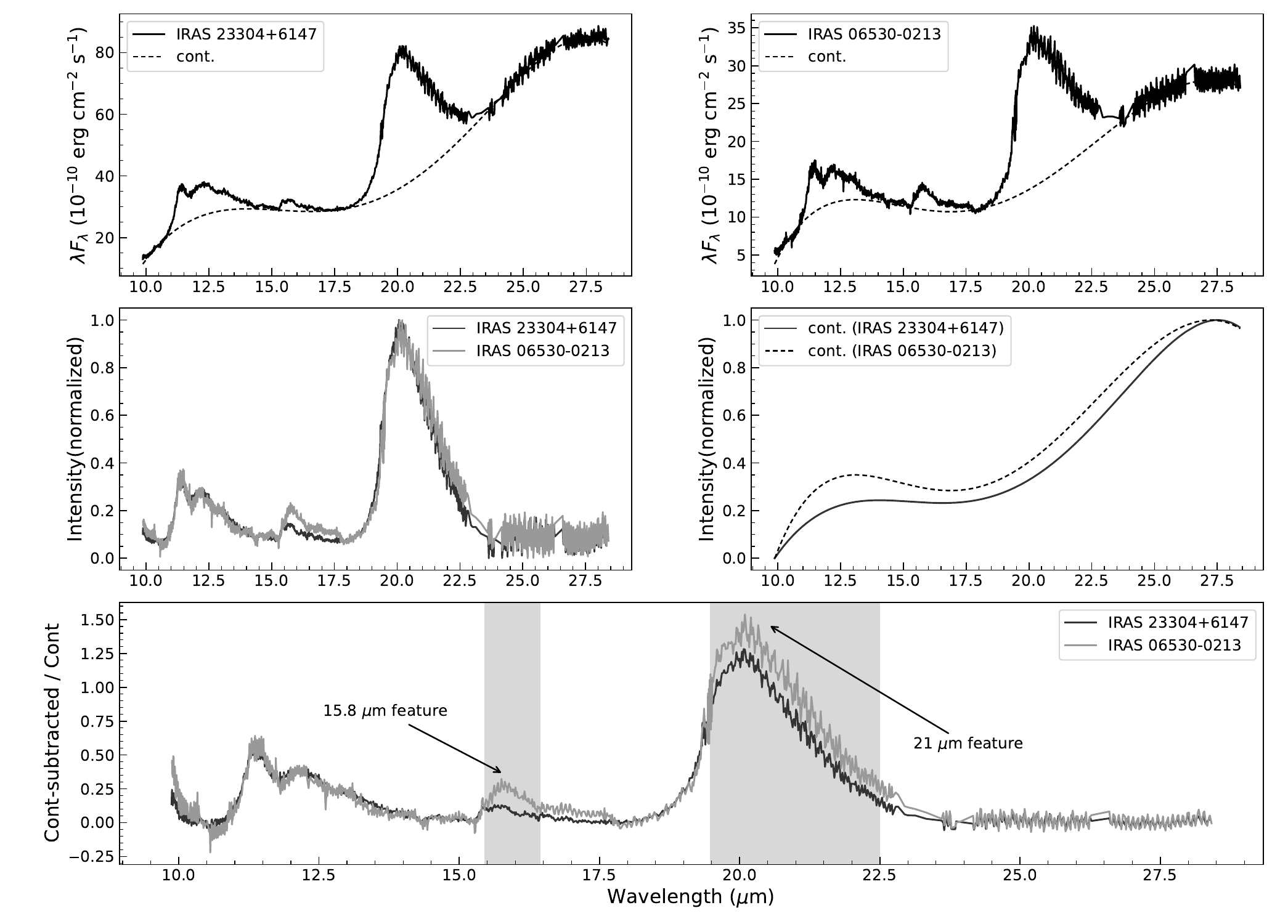}
\caption{
{\it Upper:} Spitzer/IRS spectra along with fitted polynomial continua of IRAS~06530$-$0213 and IRAS~23304$+$6147.
{\it Middle:} Normalized spectra after continuum subtraction (left panel) and normalized continuum profiles (right panel).
{\it Bottom:} Continuum-subtracted spectra divided by their continua. The shaded regions mark the 15.8~$\mu$m and 21~$\mu$m features.}
\label{infrared}
\end{figure*}

Another  21~$\mu$m source, IRAS\,23304$+$6147, also harbours an EDE structure \citep{2025A&A...696A.102S}.
Given that all imaged 21~$\mu$m  sources host EDE structures allowing complex molecules to largely survive photodissociation, we hypothesize these environments strongly influence the infrared properties of these PPNs.
Figure~\ref{infrared} compares the Spitzer/Infrared Spectrograph (IRS) spectra of the two 21\(\mu\)m sources, which were retrieved from \citet{2010ApJ...725..990Z}. Both sources display prominent emission bands at 11.2, 15.8, and 21\(\mu\)m. As indicated by the gray-shaded regions, the 15.8 and 21\(\mu\)m features are distinctly stronger in IRAS~06530$-$0213 than in IRAS~23304$+$6147, whereas their spectral profiles across the 10--15\(\mu\)m range are nearly identical. Given that IRAS~06530$-$0213 has a stronger normalized continuum across the bandpass 
and the two sources possess comparable dust temperatures \citep{2010ApJ...725..990Z}, this intensity discrepancy cannot be 
attributed to uncertainties in continuum subtraction. The enhanced 15.8 and 21\(\mu\)m features in IRAS~06530$-$0213 thus most likely arise from intrinsically stronger emission.

The 15.8\,$\mu$m emission feature is attributed to the skeletal C–C–C bending modes of large, neutral PAHs \citep{2012ApJ...747...44P,2015ApJ...811..153S}, while the carrier responsible for the 21\,$\mu$m feature has similarly been hypothesized to consist of complex organic molecules \citep{2020Ap&SS.365...88V}. Accordingly, the relatively enhanced 15.8 and 21~$\mu$m features in IRAS~06530$-$0213 may reflect more active processing, growth, or preservation of large carbonaceous/complex organic material in IRAS~06530$-$0213. 
Alternatively, the carriers of these infrared features
could have already formed in the AGB phase and become excited at the PPN stage, with IRAS~06530$-$0213 offering more favourable  conditions for their efficient excitation.

 Figure~\ref{infrared} shows that the 15.8~$\mu$m feature exhibits an asymmetric, non-Lorentzian profile. 
This band rises steeply on the blue side and features a broad, extended red wing. Compared with IRAS~23304$+$6147, IRAS~06530$-$0213 shows an even broader and flatter red wing.
Such red-wing broadening of PAH bands can be qualitatively explained by anharmonic effects. In an anharmonic oscillator, transitions originating from vibrationally excited states take place at slightly lower frequencies, or equivalently longer wavelengths, than the fundamental transition. A stronger contribution from these redshifted hot-band components thus broadens and flattens the red wing of the observed profile. This suggests that IRAS~06530$-$0213 hosts a higher temperature environment favourable for exciting these bands.
This interpretation agrees well with the stellar properties of the two sources. The central star of IRAS~06530$-$0213 is of spectral type F5~I with an effective temperature of about 6900~K, while IRAS~23304$+$6147 has a cooler central star of spectral type G2~Ia and \(T_{\rm eff}\approx5900\)~K \citep{2003ApJ...590.1049H,2000AstL...26...88K}. 
Nevertheless, the asymmetric profile may also result from blending of closely related molecular species \citep{2002A&A...388..639P,2008ARA&A..46..289T,2010A&A...511A..32B}. By contrast, the 21~$\mu$m profiles of the two sources show far greater similarity. Previous work has demonstrated that this 21~$\mu$m feature exhibits a highly uniform morphology across different PPNs \citep{2020Ap&SS.365...88V}. 
Such disparate behaviours between the 15.8~$\mu$m and 21~$\mu$m bands may imply that their respective excitation conditions 
 or chemical complexity respond divergently to variations in the local environment. 
Statistically meaningful conclusions can only be drawn once future observations of an expanded source sample become available.

\section{Conclusion} \label{sec:concl}

We present high-angular-resolution NOEMA observations of the carbon-rich 21~$\mu$m PPN IRAS~06530$-$0213 in the 1.3~mm band. 
Several molecular transitions from HC\(_3\)N, C\(_4\)H, and SiC\(_2\) are newly detected and spatially resolved in this source; we jointly analyze them with previously published CO and \(^{13}\)CO emission data. The multi-molecular maps show that different species trace different regions within the same barrel-like molecular structure. 
CO and $^{13}$CO mainly delineate the more extended barrel-like nebula, whereas HC$_3$N and C$_4$H are preferentially concentrated toward the inner and lower-latitude regions, corresponding to the inner EDE.

We carried out one-dimensional radiative transfer calculations to investigate the differing molecular distributions. Excitation conditions and optical depth alone fail to account for the inward concentration of HC\(_3\)N relative to CO, which implies its abundance is necessarily elevated within the inner EDE. This indicates that the dense inner EDE is well suited for carbon-chain molecules to form and persist. Relative to IRAS~23304$+$6147 (another known 21~$\mu$m source), IRAS~06530$-$0213 displays stronger 15.8~$\mu$m and 21~$\mu$m emission and a broader red wing for the 15.8~$\mu$m band. We conclude that its environment features higher chemical and/or radiative activity. 
Observations of a large sample of 21~$\mu$m sources are required to determine whether the EDE structure constitutes a prerequisite for the production of the 21~$\mu$m feature carrier.
The upcoming Square Kilometre Array (SKA) will provide valuable sensitivity for detecting faint molecular tracers and may offer further constraints on the molecular environments associated with the 21~$\mu$m feature.

\begin{acknowledgments}

We are grateful to the anonymous referee for careful reading of the manuscript and valuable comments that greatly improved this work.
The financial supports of this work are from the National SKA Program of China (Grant No. 2025SKA0120100), the National Natural Science Foundation of China (NSFC, No.\,12473027 and 12333005), and the Guangdong Basic and Applied Basic Research Funding (No.\,2024A1515010798). This work is based on observations carried out with the IRAM NOEMA Interferometer. IRAM is supported by INSU/CNRS (France), MPG (Germany) and IGN (Spain). 
\end{acknowledgments}

\bibliography{sample631}{}
\bibliographystyle{aasjournal}

\newpage

\begin{deluxetable}{lcccccccccc}
\tablecaption{Molecular Line Detections in IRAS~06530$-$0213
\label{tab:example}}
\tablehead{
\colhead{Molecule ID} & \colhead{Transition} & 
\colhead{Frequency } & \colhead{$I_{\mathrm{peak}}^a$} &
\colhead{$\int I\,dv$ } & \colhead{rms} & \colhead{Config.} \\
& & \colhead{(MHz)} & \colhead{(K)} & \colhead{($\mathrm{K\,km\,s}^{-1}$)} & \colhead{(K)} &
}
\startdata
SiC$_2$    & J$_{Ka,Kc}=9_{4,6}-8_{4,5}$ & 213208.03 & 0.28 & 1.41 & 0.21 & 12A  \\
           & J$_{Ka,Kc}=9_{4,5}-8_{4,4}$ & 213292.33 & 0.22 & 0.72 & 0.21 & 12A \\ 
HC$_3$N    & $J=24-23$ & 218324.71 & 0.76 & 4.99   & 0.17 & Combined \\
           & $J=26-25$ & 236512.78 & 0.27 & 1.97   & 0.05 & 10D  \\ 
C$_4$H     & $J=47/2-45/2$ & 218836.98 & 0.20 & 1.30 & 0.17 & Combined \\
           & $J=45/2-43/2$ & 218875.35 & 0.26 & 1.68 & 0.17 & Combined \\
\enddata
\tablenotetext{a}{At 210--230 GHz, the flux-to-temperature conversion factor ranges from 100 to 116 K/(Jy\,beam$^{-1}$) for the 12A configuration data, 6 to 10 K/(Jy\,beam$^{-1}$) for the 10D configuration data, and 77 to 86 K/(Jy\,beam$^{-1}$) for the combined dataset.}
\label{line}
\end{deluxetable}

\begin{deluxetable*}{l c c c c}
\tablecaption{Fractional abundances of molecules in 
IRAS~06530$-$0213 and comparison sources\label{tab:tex_ncol}}
\tablehead{
\colhead{Source} &
\colhead{$^{13}$CO} &
\colhead{SiC$_2$} &
\colhead{HC$_3$N} &
\colhead{C$_4$H}
}
\startdata
\multicolumn{5}{l}{\textbf{21~$\mu$m source}} \\
IRAS~06530$-$0213
& $(2.6\pm0.6)\times10^{-5}$
& $(2.8\pm1.4)\times10^{-8}$
& $(3.2\pm0.9)\times10^{-6}$
& $(9.4\pm0.7)\times10^{-6}$ \\
IRAS~23304+6147\tablenotemark{a}
& $1.1\times10^{-5}$
& $<2.0\times10^{-8}$
& $2.3\times10^{-8}$
& \ldots \\
IRAS~22272+5435\tablenotemark{a}
& $1.7\times10^{-5}$
& $7.5\times10^{-9}$
& $2.1\times10^{-8}$
& \ldots \\
\multicolumn{5}{l}{\textbf{Non 21~$\mu$m source}} \\
IRC+10216\tablenotemark{b}
& \ldots 
& $6.5\times10^{-7}$
& $4.1\times10^{-7}$
& $3.1\times10^{-6}$ \\
IRAS~Z02229+6208\tablenotemark{a}
& $2.0\times10^{-5}$
& $<1.3\times10^{-8}$
& $1.5\times10^{-8}$
& \ldots \\
CRL~2688\tablenotemark{c}
& \ldots
& $1.1\times10^{-7}$
& $1.2\times10^{-7}$
& $1.4\times10^{-6}$ \\
\enddata
\tablenotetext{a}{Taken from \citet{2024AJ....167...91Q}.}
\tablenotetext{b}{Taken from \citet{2025AA...699A.216A}.}
\tablenotetext{c}{Taken from \citet{2022ApJS..259...56Q}.}
\end{deluxetable*}

\end{document}